\documentstyle[12pt]{article}
\catcode`@=11
\def\NPB{{\it Nucl. Phys. }{\bf B}}
\def\PLB{{\it Phys. Lett. }{\bf B}}
\def\PRL{{\it Phys. Rev. Lett. }}
\def\PRD{{\it Phys. Rev. }{\bf D}}

\def\IJMPA{{\it Int. J. Mod. Phys. }{\bf A}}

\def\ie{{\it i.e.}}

\def\tphi{\tilde\phi}
\def\tpsi{\tilde\psi}
\let\a=\alpha
\def\ta{\tilde{a}}
\def\tA{\tilde{A}}

\def\tb{\tilde{b}}
\def\tB{\tilde{B}}

\def\rd{{\rm d}}
\def\define{\buildrel{\rm def}\over=}
\let\f=\phi

\let\g=\gamma
\def\inv#1{{\textstyle{1\over#1}}}

\def\Lb{\Lambda_b}
\let\q=\theta
\let\p=\pi
\def\Seff{S_{\rm eff}}
\let\t=\tau
\let\vd=\partial
\def\vev#1{\left\langle#1\right\rangle}
\def\tZ{\tilde{Z}}
\let\w=\omega
\def\frc#1#2{{\textstyle{#1\over#2}}}
\def\IR{\relax\leavevmode{\rm I\kern-.18em R}}
\def\ZZ{\relax\leavevmode
                         \ifmmode\mathchoice
                         {\hbox{\sf Z\kern-.4em Z}}
                         {\hbox{\sf Z\kern-.4em Z}}
                         {\lower.9pt\hbox{\scriptsize\sf Z\kern-.36em Z}}
                         {\lower1.2pt\hbox{\tiny\sf Z\kern-.36em Z}}
                          \else{\sf Z\kern-.4em Z}\fi}
\def\RR{\relax\leavevmode
                         \ifmmode\mathchoice
                         {\hbox{\sf R\kern-.4em R}}
                         {\hbox{\sf R\kern-.4em R}}
                         {\lower.9pt\hbox{\scriptsize\sf R\kern-.36em R}}
                         {\lower1.2pt\hbox{\tiny\sf R\kern-.36em R}}
                          \else{\sf R\kern-.4em R}\fi}

\def\Label#1{\label{#1}%
               \smash{\hbox to0pt{\raise1ex\hbox{\tiny[#1]}\hss}}}
\def\noLabels{\let\Label=\label}
\def\Eq#1{Eq.~(\ref{#1})}

\thicklines     

\reversemarginpar
\noLabels 
\catcode`@=12
\begin{document}

\begin{titlepage}
\begin{flushright}
CITUSC/01-47\\
VPI-IPPAP-01-05\\
hep-th/0112079\\
\end{flushright}

\begin{center}

{\large\bf
                  {de Sitter Spacetimes from
                   Warped Compactifications of IIB String Theory}}\\[5mm]
{\bf P. Berglund\footnote{e-mail: berglund@citusc.usc.edu} } \\[1mm]
                   CIT-USC Center for Theoretical Physics\\
                   Department of Physics and Astronomy\\
                   University of Southern California\\
                   Los Angeles, CA 90089-0484\\[2mm]
{\bf T. H\"{u}bsch\footnote{e-mail: thubsch@howard.edu}%
               $^,$\footnote{On leave from the ``Rudjer Bo\v skovi\'c''
                   Institute, Zagreb, Croatia.} } \\[1mm]
                   Department of Physics and Astronomy\\
                   Howard University\\
                   Washington, DC 20059\\[2mm]
{\bf D. Minic\footnote{e-mail: dminic@vt.edu} } \\[1mm]
		          Institute for Particle Physics and Astrophysics\\
                   Department of Physics\\
                   Virginia Tech\\
                   Blacksburg, VA 24061\\[5mm]

{\bf ABSTRACT}\\[3mm]
\parbox{4.9in}{We continue our study of codimension two solutions of warped
space-time varying compactifications of string theory. In this letter
we discuss a non-supersymmetric solution of the classical type IIB
string theory
with de Sitter gravity on
a codimension two uncompactified part of spacetime.
A non-zero positive value of the cosmological constant is induced by
the presence of
non-trivial stringy moduli, such as the axion-dilaton system for the
type IIB
string theory.
Furthermore, the naked singularity of the codimension two solution is
resolved by the presence of a small but non-zero cosmological constant.}
\end{center}
\end{titlepage}

The issue whether de Sitter  space can be obtained from string theory
has recently attracted renewed attention~\cite{dsrev}.
In this letter we address this question by examining non-singular
non-static spacetimes
which fall into the class of codimension two non-supersymmetric
solutions of string
theory studied in Refs.~\cite{bhm1, bhm3, bhm4}.
We construct classical non-supersymmetric string theory
solutions in $D$ dimensions in which the metric in the uncompactified
$(D-2)$ directions
is de Sitter rather than flat Minkowski space as was done
in Refs.~\cite{bhm1, bhm4}. The latter theories have a naked singularity
in analogy with the
global cosmic string~\cite{cohen}. The main point of this letter is that,
in non-supersymmetric codimension two solutions of string theory (either
type IIB or any space-time variable string vacua),
a positive cosmological
constant, $\Lb$, in the $D{-}2$ dimensional cosmic brane spacetime
naturally resolves the naked singularities in the transverse 2-plane\footnote{
Note that $\Lb>0$, {\it removes} the naked singularity present in the model considered in~\cite{bhm1, bhm3, bhm4}, in comparison with earlier discussions of a positive cosmological constant along the brane-world~\cite{kachru}.}.
We show how the
naked singularity of the global cosmic brane configuration is deformed to a
cosmological horizon while keeping the overall features of the solution
away from the horizon\footnote{Gregory was first to point out the
analogous observation in the context of global
$U(1)$ vortex solutions~\cite{RG,RGII}: the naked singularity of the
global cosmic string~\cite{cohen} is resolved into a non-singular
non-static spacetime.}.  In fact, it is the stringy moduli, with
non-trivial
$SL(2,\ZZ)$ transformation properties, that lead to $\Lb>0$.
Thus, string theory gives rise to a smooth background
with a positive cosmological constant in the effective $D-2$ dimensional
theory.

The analysis concerning the existence of a normalizable zero mode, and hence
gravity, in the compactified $D-2$ dimensional theory follows
our previous work~\cite{bhm4}. The higher KK-modes and the
corrections to the Newton potential are suppressed, due to
the presence of nontrivial moduli. Of course, this
is intuitively expected from the viewpoint of generic Kaluza-Klein
compactification.  Our analysis is closely related to the
situation without matter fields in which case one obtains a Rindler-space
type solution~\cite{kaloper} with non-trivial boundary conditions. In the
latter case if one assumes a realistic value of the cosmological constant,
the KK modes give a very large correction to the Newton potential and
the effect of the lower dimensional graviton is washed out~\cite{kaloper}.

Let us contrast the above discussion with the Karch and Randall
solutions~\cite{rk1}
in which the bulk cosmological constant {\it and} the brane cosmological
constant are both
negative. These solutions describe localized gravity on anti-de Sitter (AdS) 
spaces which are 
natural from the point of view of supersymmetry.
Note that Gregory has shown that the solutions with the negative
brane cosmological constant appear as repulsive fixed points in the
space of warped factors describing the codimension two solutions
of the global vortex type~\cite{RGII}.
We indeed find that
the de Sitter (or accelerating universe) like solutions appear
to be more generic and natural in the space of all non-supersymmetric
solutions of the IIB string theory.

Let us briefly review the general framework of global cosmic branes from a
string theory perspective. We consider a higher-dimensional string theory
compactified on a Calabi-Yau (complex) $n$-fold, some moduli
of which are allowed to vary over (the `transversal') part of the non-compact
space~\cite{vafa,gh}. Specifically, we will consider type IIB string
theory (possibly compactified on a {\it fixed} manifold) in which we let
the axion-dilaton system, $(\a,\f)$, described by the complex modulus field
$\t=\a + i\exp(-\phi)$,
vary over the uncompactified spacetime.  Following
Ref.~\cite{bhm1}, the effective action describing the coupling of the
moduli to gravity of the observable spacetime can be derived by
dimensionally reducing the higher dimensional Einstein-Hilbert
action~\cite{vafa,gh}. Thus, the relevant part of the low-energy effective
$D$-dimensional action of the modulus, $\t$, of the Calabi-Yau
$n$-fold coupled to gravity reads
\begin{equation}
            \Seff
            = {1\over2\kappa^2}\int\rd^D x \sqrt{-g} ( R
               - {\cal G}_{\t \bar{\t}}g^{\mu \nu}
                 \vd_{\mu} \t \vd_{\nu} \bar{\t}
                      +...)~.
\Label{e:effaction}
\end{equation}
Here $\mu,\nu=0,{\cdots},D-1$, $2\kappa^2=16\p G^D_N$, where $G_N^D$
is the $D$-dimensional Newton constant, and ${\cal
G}_{\t\bar\t}=-(\t-\bar \t)^{-2}$ is the metric on ${\cal M}$,
the complex structure moduli space of a torus. We neglect higher
derivative terms in this effective action and set the other fields in the
theory to zero as in Ref.~\cite{vafa}. We also restrict the modulus to
depend on $x_i$, $i{=}D{-}2,D{-}1$, so that $\vd_a \t{=}0$,
$a{=}0,{\cdots},D{-}3$.

With the above setting of the $D{=}10$ type~IIB string theory, we find that
the explicit solutions for $\t$ are aperiodic, but do exhibit a non-trivial
$SL(2,\ZZ)$ monodromy~\cite{bhm1}, which ensures this to be a
stringy (although classical and non-supersymmetric) rather than merely
a supergravity vacuum. Being a modulus, $\t$ interacts with
other fields only through gravity.  Therefore, although supersymmetry
will turn out to be broken (see also Ref.~\cite{bhm1}), the induced
potential for $\t$ can be safely neglected.

The absence of a potential for $\t$ permits the following
simplification of the Einstein equation:
\begin{equation}
     R_{\mu\nu} ~=~ {\cal G}_{\t\bar{\t}}\,\vd_{\mu} \t\vd_{\nu}\bar\t
     \define\widetilde{T}_{\mu\nu}~.
\Label{e:EinStein}
\end{equation}
\Eq{e:EinStein} affords the `separation of variables' where the metric is
axially symmetric, while
$\t$ is independent of the radial distance from the cosmic brane, so
$\widetilde{T}_{\mu\nu}=\hbox{diag}[0,{\cdots},0,\inv4\w^2l^{-2}]$.
(From Ref.~\cite{bhm1}, $\t=\alpha_0+i g_s^{-1}\exp(\omega \q)$ and in
particular the dilaton of the type~IIB superstring theory varies with the
polar angle, not the radial distance.)
\Eq{e:EinStein} then defines the general class of our spacetimes as {\it
almost\/} Ricci-flat: $R_{\mu\nu}=\hbox{diag}[0,{\cdots},0,\inv4\w^2
l^{-2}]$, where $\w^2{>}0$ is indeed related to supersymmetry
breaking~\cite{bhm1} and $l$ is the is length scale set by the global
cosmic brane.

With a phenomenologically interesting $K3$ compactification of
the $D{=}10$ solution in mind (upon which the metric receives $\alpha'$
corrections), we continue with the general $D$-dimensional setting. The
Ansatz for the metric, with $z=\log(r/l)$,
is given by:
\begin{eqnarray}
       \rd s^2 &=& A^2(z)\, \bar g_{ab}\rd x^a \rd x^b
                 + l^2 B^2(z)\,(\rd z^2 + \rd\q^2)~, \Label{e:Metric} \\[1mm]
       \bar g_{ab}\rd x^a \rd x^b &=& - \rd x_0^2 +
            e^{2\sqrt{\Lb} x_0}\,(\rd x_1^2 + \ldots + \rd x_{D-3}^2)~,
                 \Label{e:gbar}
\end{eqnarray}
Incidentally, it is easy to prove that if $\bar{g}_{ab}$ is chosen
to be any Ricci-flat metric (e.g., the Schwarzschield geometry), the
solutions of Refs.~\cite{bhm1,bhm4} remain unchanged. This can be
done in complete analogy with Refs.~\cite{real, myers}. Hereafter, however,
we focus on the de~Sitter metric~(\ref{e:gbar}).

The $R_{ab}=0$ part of \Eq{e:EinStein} reduces
to a single equation, giving:
\begin{equation}\small
     B^2 =  l^{-2}\Lb^{-1}\Big(A'{}^2+{1\over(D{-}3)} A A''\Big)=
l^{-2}\Lb^{-1} \frac{h''  h^{-\frac{D-4}{D-2}}}{(D-2)} ~,
\Label{e:BfromA}
\end{equation}
which determines $B(z)$ in terms of $A(z)$ or $h(z)\define A(z)^{D-2}$.
With this substitution, the remaining components of
\Eq{e:EinStein} produce the following single equation~\footnote{It is
straightforward to show that $R_{zz}$ and $R_{\q\q}$ can be written as
certain linear combinations of the above differential equation and its
derivatives.}:
\begin{equation}
      \frac{1}{2(D-2)}\frac{h'{}^2}{h^2} - \frac{h''}{2h}+\frac{h'h'''}{2h
h''}= -\frac{1}{8}\omega^2\,.~~~~ \Label{e:h}
\end{equation}
This implies that $\Lb{>}0$, and that the
Ansatz~(\ref{e:Metric})--(\ref{e:gbar}) does not permit a double Wick
rotation into an AdS spacetime. To see this, note
that \Eq{e:h} determines $h(z)$, and hence $A(z)$, to be independent of
$\Lb$. But then, $\Lb\to-\Lb$ in \Eq{e:BfromA} would imply $B(z)^2<0$,
making the entire plane transverse to the cosmic brane also time-like.

Furthermore, with $h(z)=(1- z/\rho_0)^{D-2}$, and so with
\begin{equation}
        A_0(z)=Z(z)\define(1- z/\rho_0)\,,\quad\hbox{and}\quad
        B_0(z)\define{1\over l\rho_0\sqrt{\Lb}}~,
\Label{e:ABzero}
\end{equation}
the metric~(\ref{e:Metric}) satisfies the Einstein
equations~(\ref{e:EinStein}) for $\w^2=0$, \ie, when $\t=\hbox{\it
const.}$ This solution describes the familiar Rindler
space~\cite{kaloper}.

For $\w\neq0$ ($\t\neq\hbox{\it const.}$), \Eq{e:h} has a
perturbative solution\footnote{This solution is of the same form as
that discussed by Gregory~\cite{RG,RGII} for the $U(1)$ vortex solution.}:
\begin{eqnarray}
A(z) &=& Z(z) \Big(1- {\w^2 \rho_0^2(D-3)\over 24
(D-1)(D-2)} Z(z)^2 + O(\w^4)\Big)~,
\Label{e:A}\cr
B(z) &=& {1\over l\rho_0\sqrt{\Lb}}\Big(1 -  {\w^2\rho_0^2\over
       8(D-1)} Z(z)^2 + O(\w^4)\Big)~. \Label{e:B}
\end{eqnarray}
Notice that, as the metric depends on $A(z)^2$ and $B(z)^2$, it is
well-defined for all values of $z$. In particular, $z \sim \rho_0$ is the
location of a putative horizon~\cite{RG,RGII}. It is easy to check for our
solution~(\ref{e:B}) that both the Ricci scalar and tensor vanish at
$z=\rho_0$, as does the whole Riemann tensor.
In fact, these tensors as well as the $R_{\mu\nu}R^{\mu\nu}$ and
$R_{\mu\nu\rho\sigma} R^{\mu\nu\rho\sigma}$
     curvature scalars all remain bounded away from $z=\rho_0$. Thus,
close to the horizon spacetime is asymptotically flat in agreement with
the behavior of Rindler space,
see Eqs.~(\ref{e:ABzero})--(\ref{e:B})~\cite{kaloper}.

In contrast, when $\Lb=0$ the solution is very different~\cite{bhm1},
\begin{equation}
       \tA(z) = \tZ(z)^{1\over(D-2)} ~,\qquad
       \tB(z) = \tZ(z)^{-(D-3)\over 2(D-2)}
               e^{{\xi\over 2a_0}(1-\tZ(z)^2)}~,
\Label{e:oldsolution}
\end{equation}
where now $\tZ=(1-a_0 z)$, and we restrict to $a_0>0$. This
solution exhibits a naked singularity, at $z=a_0^{-1}$ ($\tZ=0$),
for the global cosmic brane and the region $z>a_0^{-1}$ ($\tZ<0$) is
unphysical: the metric becomes complex. In the solution~(\ref{e:A}),
the singularity is effectively removed by introducing the non-zero
cosmological constant along the brane.

While the naked singularity has been removed by the non-zero $\Lb$
we will now show that away from the horizon,
the global cosmic brane solution~(\ref{e:oldsolution}) is still a good
approximation to \Eq{e:B}. We first obtain a power series solution of
Eqs.~(\ref{e:B}), expanding around $z=0$. From this we determine the
lowest order terms in $h(z)=\sum_{n=0} h_n z^n$~\footnote{This requires
an initial guess for the value of $\omega^2\rho_0^2$ and that the higher
order corrections in the expansion of $A(z)$ in terms of $Z(z)$ fall off
fast enough. Indeed, we have computed the expansion of $A(z)$ to
$O(Z^{12}(z))$, and determined $\omega^2\rho_0^2$ and the corresponding
numerical values of the coefficients $h_i$ recursively.}.
Finally, we expand $A(z)$ and $B(z)$, expressed as functions of $h(z)$
and $h''(z)$ to lowest order in $z$,
\begin{equation}
      A(z) = (1+  z \frac{h_1}{(D-2)})\,,\quad
B(z) = \sqrt{l^{-2}\Lb^{-1} \frac{2h_2}{D-2}}\Big(1+z(\frac{3
h_3}{2h_2} -\frac{h_1(D-4)}{2(D-2)})\Big)~.
\Label{e:ABexpansion}
\end{equation}
Here, the coefficients $h_i$ for $i>2$ are determined in terms of
$h_0,h_1,h_2$ by \Eq{e:h}, the overall rescaling of $A(z)$ and
$B(z)$ is absorbed in a rescaling of $x^a$ and $l$, respectively, and
the numerical values of $h_1,h_2$ are determined by comparison with the
expansions~(\ref{e:B}).
Comparing now \Eq{e:ABexpansion} with \Eq{e:oldsolution}, expanded
to first order in $z$, leads to
\begin{equation}
   a_0\approx 0.9\,(D{-}2)\rho_0^{-1}~,\quad
   \xi\approx\inv{0.9}\,\w^2\rho_0^2 \frac{\rho_0^{-1}}{8(D-2)}~,
    \quad\hbox{and}\quad
   l^{-2}\Lb^{-1} \frac{\omega^2}{2(D-2)}=1~.  \Label{e:OldIsNew}
\end{equation}

The last of the identifications~(\ref{e:OldIsNew}) has the following
important consequence:
\begin{equation}
      \Lb=\frac{\omega^2}{2(D-2)l^2}~.
\Label{e:dSomega}
\end{equation}
Thus, the cosmological constant is directly related to the amount of
matter, or rather the non-trivial variation of the matter as a function of
$\theta$! This gives a very non-trivial relation between the stringy moduli,
and hence string theory itself, and a positive $\Lb$.
Since the dilaton is
$\phi=-\omega\q$ it also follows from \Eq{e:dSomega} that we have a
strongly coupled theory~\footnote{Recall that with $\t=\alpha_0 + i
g_s^{-1}\exp(\omega \q)$, the $SL(2,\ZZ)$ symmetry requires $g^D_s\sim
O(1)$ in $D$ dimensions. However, in the $D{-}2$-dimensional
brane-world, $g_s^{D-2}=g_s^D\sqrt{\a'/V_\perp}$, and since $V_\perp$, the 
volume of the transverse space, is  large~\cite{bhm1}, $g_s^{D-1}\ll1$.}.

Note also that  $\Lb \sim \omega^2/l^2$ is consistent
with the notion that supersymmetry
breaking and a non-zero cosmological constant are related. In our
scenario supersymmetry is explicitly broken by $\omega^2 \neq 0$. But
since $\Lb \sim \omega^2/l^2$, supersymmetry breaking by
$\omega^2 \neq0$ also induces a positive cosmological constant, which
then can vanish only in the decompactifying limit, $l\to\infty$.
In the limit
$\omega^2 =0$ we recover supersymmetry and thus have a possible
F-theory~\cite{FTh} background.

     From the previous analysis we know the metric in two different regions:
(far) away from the horizon~(\ref{e:oldsolution}), and near the
horizon~(\ref{e:ABzero}). Thus we can determine Kaluza-Klein corrections
to the Newton potential.
We start by analyzing the small gravitational fluctuations
$\delta\eta_{ab}=h_{ab}$ (far) away from the horizon. This
analysis follows to a large degree the discussion in Refs.~\cite{bhm1,bhm4}.
{}From the Einstein equations,
$h_{ab}$ satisfies a wave equation of the form~\cite{csaki}:
\begin{equation}
          \Box h_{ab} = \frac{1}{\sqrt{-g}}\,\vd_{\mu}(\sqrt{-g}
                        g^{\mu\nu}\vd_{\nu} h_{ab}) = 0~. \Label{e:WEq}
\end{equation}
Following Ref.~\cite{cohen} we  change coordinates, with $\tA,\tB$ given in
\Eq{e:oldsolution},
\begin{eqnarray}
   \rd v &=& l\, \frac{\tB}{\tA}\rd z~
\Label{e:dvdz} \\[1mm]
   \rd s^2 &=& \tA^2(v) \eta_{ab}\rd x^a \rd x^b
              + \tA^2(v) \rd v^2 + \tB^2(v) l^2 \rd\q^2~. \Label{e:vMetric}
\end{eqnarray}
The isometries of the metric dictate the following Ansatz
\begin{equation}
           h_{ab}=\epsilon_{ab} e^{ip\cdot x} e^{in\q}
\frac{\tphi_m}{\tpsi_0}~,\quad{\rm where}\quad
             \tpsi_0\define\sqrt{\tA^{-2}\sqrt{-g}}
             =\sqrt{\tA^{D-3} \tB}~.
\Label{e:psizero}
\end{equation}
With these variables~\cite{bhm1, cohen}, \Eq{e:WEq}
becomes a Schr\"{o}dinger-like equation:
\begin{equation}
              -\tphi_m^{''} + \Big(\frac{\tpsi_0^{''}}{\tpsi_0} +
\frac{\tA}{\tB}n^2\Big)\tphi_m
               = m^2\tphi_m~. \Label{e:Schrodinger}
\end{equation}
For simplicity, we choose $n=0$.
Integrating \Eq{e:dvdz} gives
\begin{equation}
             v=v_*\Big[1-\g_0^{-1}
              \g\Big(\frc{D-3}{4(D-2)};\frc{\xi Z^2}{2a_0}\Big)
               \Big]~, \quad
             v_*=\frc{l}{2a_0}\,e^{\xi\over 2 a_0}
             \Big(\frc{2a_0}{\xi}\Big)^{{D-3\over4(D-2)}}\,\g_0~,\quad
             \g_0= \g\Big(\frc{D-3}{4(D-2)};\frc{\xi}{2a_0}\Big)~.
             \Label{e:VzChng}
\end{equation}
where $\gamma$ is the incomplete $\Gamma$-function.
Before turning on the brane cosmological constant
the naked singularity is situated at $z=1/a_0$ or equivalently at
$v=v_*$. Since the naked singularity is effectively removed when
$\Lb>0$ the analysis below is only valid for
$v\in[0,v_*-\delta v_*]$, where $v_*-\delta v_*$
is the location at which the true metric starts to substantially
deviate from the global cosmic brane metric\footnote{We regard the
global cosmic brane metric~(\ref{e:oldsolution}) to break down when
$h_0 + h_1 z = \sum_{n=2}^\infty h_n z^n$, \ie\ when the sub-leading
order corrections start to dominate.}. One can show that
$\delta v_*/v_* \approx 0.5$.
When $v>v_*-\delta v_*$ we have to use the form of the metric given
by the de Sitter  perturbed solution~(\ref{e:ABzero}).

With the above change of variables the zero-mode becomes
$\tpsi_0(v)=\sqrt{l\,(1-v/v_*)}$ in the $\xi=0$ limit.
With this $\tpsi_0$ the general solution of \Eq{e:Schrodinger} is
\begin{equation}
\tphi_m = \ta_m \sqrt{v_*-v}\,J_0\Big(m(v_*-v)\Big) +
                     \tb_m \sqrt{v_* - v}\,Y_0\Big(m(v_*-v)\Big)~.
                     \Label{e:Bessel}
\end{equation}
When $\Lb=0$ it is possible to choose boundary conditions
such that $\tb_m=0$. However, one can show that as we approach
$v_*-\delta v_*$, a non-zero $\tb_m$ is not be consistent with the
appropriate boundary conditions at the location of the horizon.

We now look at small gravitational
fluctuations $\delta\eta_{ab}=h_{ab}$
of the longitudinal part of the metric close to the cosmological horizon.
{}From the Einstein equations,
$h_{ab}$ satisfies a wave equation of the form given in
\Eq{e:WEq}. The isometries of the metric dictate the following
Ansatz
\begin{equation}
           h_{ab}=\epsilon_{ab} \varphi_p(x) e^{in\q	}
\frac{\phi_m}{\psi_0}~,
\end{equation}
where $\varphi_p(x)$ is a mode with momentum $p$ along the de Sitter space,
with coordinates $x$,
\ie\
\begin{equation}
       (\Box_x + 2\Lb)\varphi_p(x) = m^2 \varphi_p(x)~,
\end{equation}
and $\psi_0$ is defined as in \Eq{e:psizero}
with $A,B$ given in Eqs.~(\ref{e:B}).
The resulting differential equation is of the same form as
in \Eq{e:Schrodinger}.

      As in the analysis above, we focus on the $s$-wave, $n=0$. We
start by considering the case in which we can
ignore the higher order corrections to the metric~(\ref{e:B}), or
equivalently, ignore the matter contribution, \ie\
$\w^2=0$. It is then possible to explicitly invert the
relation~(\ref{e:dvdz}), so
   $e^{-k v} = (1-\rho_0^{-1} z)$ where  $k=\frac{1}{\rho_0 l}$.
Thus, the zero-mode is given by \Eq{e:psizero}~\footnote{Since
$\psi_0$ is only valid for $v\in[v_*-\delta v_*,\infty)$, we have to
match with $\tpsi_0$ at $v\sim v_*-\delta v_*$.}
\begin{equation}
       \psi_0(v) = (l\frac{\delta v_*}{v_*})^{1/2}e^{\frac{(D-3)}{2} k
(v_*-\delta v_*-v)}~.
\end{equation}
The solutions to \Eq{e:Schrodinger} are then
given as
\begin{equation}
\phi_m(v) = a_m  e^{-\sqrt{\frac{(D-3)^2k^2}{4} - m^2}\,v} +
                     b_m e^{\sqrt{\frac{(D-3)^2k^2}{4} - m^2}\,v}~.
                     \Label{e:Planewave}
\end{equation}
When $m^2<(D-3)^2k^2/4$ the normalizability of the modes implies that
$b_m=0$. When $m^2>(D-3)^2k^2/4$ the argument of the exponential is
$\pm i\sqrt{\frac{-(D-3)^2k^2}{4} +m^2}\,v$, \ie\
\begin{equation}
       \phi_m(v) = \frac{a_m}{m^{1/2}}
                    \cos\Big(\sqrt{-\frac{(D-3)^2k^2}{4}+ m^2}\,v\Big)~.
                     \Label{e:Planewavelargev}
\end{equation}
having again matched the wave functions at $v=v_*-\delta v_*$;
hereafter we drop the tildes.

          We are left to determine $a_m$ such that
$\vev{\phi_m|\phi_m}=1$.
Because $J_{0}^2(mv_*) \sim \frac{\cos^2(mv_* -\pi/4)}{mv_*}$
when $mv_*\gg1$, we can rewrite the normalization integral as
\begin{equation} 1~=~
  \vev{\phi_m|\phi_m}~\sim~\frac{a_m^2}{m}\int_{0}^\infty\rd v~\cos^2(mv)
  ~\sim~ a_m^2 \frac{v_c}{m}~,
\end{equation}
where we have regularized the integral by putting the system in a box of
size
$v_c$. Therefore, $a_m\sim m^{1/2}(v_c)^{-1/2}$.

The last step is to include the contribution of the matter, $\xi>0$. This
will only effect the wavefunction (far) away from the horizon, since
close to the horizon the metric becomes independent of $\xi$.
However, since $a_0\sim \xi$ the
ratio $\xi/a_0$ can be neglected in the definition of $v_*$, see
\Eq{e:VzChng}. Thus, the effect of  $\xi>0$ is negligible.

Similarly, for the zero-mode normalization, or equivalently the volume of the transverse space $V_\perp$, we get
\begin{equation}
    \vev{\psi_0|\psi_0}\sim
     \pi l v_* \Big[1-(\frac{\delta v_*}{v_*})^2\Big] +
      2 \pi v_* l \frac{2}{D-3} (\frac{\delta v_*}{v_*})\frac{1}{k v_*}
  \approx \pi lv_*\Big[1-(\frac{\delta v_*}{v_*})^2
                        +2{\cdot}0.9\,(\frac{\delta v_*}{v_*})\Big]~.
\end{equation}
where we have used that $v_* k\approx 2/(0.9 (D-3))$ which follows from
the definitions of $k$ and $v_*$. Since $\delta v_*/v_*\sim 0.5$ we see
that contributions from the two different regions are of the same order.
Finally, we can rewrite the volume integral in terms of
$\Lb$ using \Eq{e:OldIsNew},
\begin{equation}
    \vev{\psi_0|\psi_0}\sim \frac{\pi}{D-3} \frac{l}{\sqrt{\Lb}}~.
\end{equation}

With these facts let us restrict our attention to $D=6$, by
compactifying the type IIB theory on a $K3$ or $T^4$ manifold.
The Newton potential is found to have the following form:
\begin{equation}
             U(r) \sim M_6^{-4} \frac{M_1 M_2}{r} \vev{\psi_0^2}
             + \sum_mM_6^{-4} \frac{M_1 M_2}{r} e^{-mr} \vev{\phi_m^2}~,
\Label{e:Newt}
\end{equation}
where $\vev{\psi_0^2}$ and $\vev{\phi_m^2}$ are the
wavefunctions averaged over the transverse space, \ie\
  $\vev{\psi_0^2}=\vev{\psi_0|\psi_0}^{-1}$ and
  $\vev{\phi_m^2}=\vev{\psi_0|\psi_0}^{-1}$.
  The Newton constant is determined from the first term:
              $G_N \sim M_6^{-4} \vev{\psi_0^2}$,
and the correction to Newton's potential is
\begin{equation}
    \triangle U(r) \sim G_N \frac{M_1M_2}{r}  \sum_m e^{-mr}~,
\end{equation}
and  can be neglected when
$r\gg r_0=(\vev{\psi_0^2})^{-1/2}\sim\Lb^{-1/4} l^{1/2}$,
where the mass gap is
$\delta m = (\vev{\psi_0^2})^{1/2}\sim \Lb^{1/4} l^{-1/2}$.
With a realistic value for
$\Lb\sim G_N  10^{-44}\, (\hbox{GeV})^4$
and taking $l=l_{\rm Pl}$, we find that $M_6\sim 10^3\, \hbox{GeV}$.
Furthermore, we get
$r_0\sim10\,\mu$m and
$\delta m\sim 10^{-2}\,$eV, which is
close to the lower bound at which gravity has been tested. Thus, the KK
modes are suppressed, due to the peculiar dependence on $\Lb$, in
contrast with the situation for Rindler space~\cite{kaloper}.

In conclusion, in this letter we have shown that the space of
non-supersymmetric codimension two solutions of string theory (either
type IIB or any space-time variable string vacua) naturally have a
positive  cosmological constant,
$\Lb$ in $D-2$ dimensions, if one {\it requires} the solution to be
non-singular.
The fact that our solutions were
obtained in classical string theory also brings some new features:
The non-zero positive value of the cosmological constant is essentially
implied by the presence of non-trivial stringy moduli, such as the
axion-dilaton system for the  Type-IIB
supergravity theory. This also implies that the string theory is
strongly coupled. The fact that the value of the cosmological constant is
directly related to the amount of matter invokes comparisons with various
attempts to incorporate Mach's principle in string theory~\cite{mach} as
well with the idea that supersymmetry breaking might have cosmological
origin~\cite{banks}.
Note that our codimension two solutions can be intuitively
understand as a strongly coupled classical non-supersymmetric solution of
F-theory~\cite{FTh}. In the $\omega =0$ limit we obtain a solution that
has similar properties to a collection of 12 $D7$-branes~\cite{bhm1}.
Thus from the metric point of view, de Sitter like (accelerating
universe) solutions appear to be
more generic and natural in the space of non-supersymmetric classical
vacua. The stability of this solutions, due to the presence of the bulk
horizon can be presumably analyzed from a purely thermodynamic  point of
view.

Unlike the Randall-Sundrum set-up~\cite{rs1, rs2}, which can be
in principle understood from the viewpoint of flux
compactifications~\cite{verlinde} in string theory, our codimension two
solutions still lack such an approach. As shown in this letter
(and as first pointed out by Gregory~\cite{RG, RGII}
in a related context) the codimension two solutions are in a very precise
sense attracted to a more general class of solutions with a positive
cosmological constant. The Randall-Sundrum scenario seems to mesh more
naturally with a negative brane cosmological constant~\cite{rk1}.
In this context a codimension one AdS type brane world
immersed in another AdS type bulk world can be realized, at least in the
situation with no gravitational back reaction, as a background seen by a
spectator $D5$ brane  in the presence of a large number of $D3$ branes.
Now, it has been argued that a certain (unstable) configurations
of type $IIB*$ theory -$E4$ brane - realizes a de Sitter geometry in
the near horizon limit~\cite{hull}.
It is tempting to try to understand
de Sitter like brane worlds from the point of view of $E$ - $D$ brane
systems. (For example a de Sitter like brane world embedded in
a bulk AdS space could be "seen"
by a spectator $E$ brane in the background of many $D$ branes.)

Finally, there are codimension one solutions of $USP(32)$ string
theory~\cite{usp} which share similar metric properties
with our codimension two solutions - in particular
the naked singularities exist in both cases at finite proper
distance from the core of the solution. In particular, it is possible
that an analysis analogous to ours would reveal
that naked singularities appearing in Ref.~\cite{usp} are also resolved
by turning on a small positive brane cosmological constant.
We hope to investigate some of these issues in the future.

{\bf Acknowledgments:}
We thank V.~Balasubramanian,  P.~Horava, N.~Kaloper and R.~Myers
for useful discussions. P.~B. would like to thank  LBL, the Berkeley Center
for Theoretical Physics and in particular the CIG, Berkeley
for  their hospitality.
                  The work of P.~B.\ was supported in part by  the US
Department of Energy under grant number DE-FG03-84ER40168.
                  T.~H.\ wishes to thank the Caltech-USC Center for Theoretical
Physics for its hospitality, and the US
Department of Energy for their generous support under grant number
DE-FG02-94ER-40854.
                  The work of D.~M.\ was supported, in the initial 
stages of this
project, by the US
Department of Energy under grant number DE-FG03-84ER40168.

\end{document}